\documentclass[12pt]{article}
\usepackage{amssymb,amsmath}
\usepackage{epsfig}
\usepackage{array}
\makeatletter
\def\alt{\mathrel{\mathpalette\gl@align<}}
\def\agt{\mathrel{\mathpalette\gl@align>}}
\def\gl@align#1#2{\lower.6ex\vbox{\baselineskip\z@skip\lineskip\z@
\ialign{$\m@th#1\hfil##\hfil$\crcr#2\crcr\sim\crcr}}} \makeatother

\def \gtsim    {\relax\ifmmode{\mathrel{\mathpalette\oversim >}}
                  \else{$\mathrel{\mathpalette\oversim >}$}\fi}
\def \ltsim    {\relax\ifmmode{\mathrel{\mathpalette\oversim <}}
                  \else{$\mathrel{\mathpalette\oversim <}$}\fi}
\def\oversim#1#2{\lower4pt\vbox{\baselineskip0pt \lineskip1.5pt
            \ialign{$\mathsurround=0pt#1\hfil##\hfil$\crcr#2\crcr\sim\crcr}}}
\newcommand{\gev}  {\mbox{${\rm GeV}$}}

\newcommand{\invfb}{\mbox{${\rm fb}^{-1}$}}

\newcommand{\faketau} {\mbox{$f_{j\rightarrow\tau}$}}
\newcommand{\pt}  {\mbox{$p_{\rm T}$}}
\newcommand{\ptvis}  {\mbox{$p_{\rm T}^{\rm vis}$}}
\newcommand{\et}  {\mbox{$E_{\rm T}$}}

\newcommand{\met} {\mbox{${E\!\!\!\!/_{\rm T}}$}}

\newcommand{\dM}{\mbox{$\Delta M$}}

\newcommand{\mtautaumax}{\mbox{$M_{\tau\tau}^{\rm max}$}}
\newcommand{\mtautauvis}{\mbox{$M_{\tau\tau}^{\rm vis}$}}
\newcommand{\mtautaupeak}{\mbox{$M_{\tau\tau}^{\rm peak}$}}

\newcommand{\bbar} {\mbox{$\overline{b}$}}
\newcommand{\tbar} {\mbox{$\overline{t}$}}

\newcommand{\ttbar}{\mbox{$t\overline{t}$}}

\newcommand{\tauh} {\mbox{$\tau_{\rm h}$}}
\newcommand{\tanb} {\mbox{$\tan\beta$}}
\newcommand{\mzero}{\mbox{$m_{0}$}}
\newcommand{\mhalf}{\mbox{$m_{1/2}$}}

\newcommand{ \gluino}   {\mbox{$\tilde{g}$}}
\newcommand{ \squark}   {\mbox{$\tilde{q}$}}

\newcommand{ \squarkR}   {\mbox{$\tilde{q}_{R}$}}

\newcommand{ \usquarkL}  {\mbox{$\tilde{u}_{L}$}}
\newcommand{ \usquarkR}  {\mbox{$\tilde{u}_{R}$}}

\newcommand{ \sbottomone}{\mbox{$\tilde{b}_{1}$}}

\newcommand{ \stopone}  {\mbox{$\tilde{t}_{1}$}}

\newcommand{ \stoptwo}  {\mbox{$\tilde{t}_{2}$}}

\newcommand{ \seleR}    {\mbox{$\tilde{e}_{R}$}}

\newcommand{ \seleL}    {\mbox{$\tilde{e}_{L}$}}

\newcommand{ \smuR}     {\mbox{$\tilde{\mu}_R$}}

\newcommand{ \stauone}  {\mbox{$\tilde{\tau}_{1}$}}
\newcommand{ \stauonep} {\mbox{$\tilde{\tau}_{1}^{+}$}}
\newcommand{ \stauonem} {\mbox{$\tilde{\tau}_{1}^{-}$}}

\newcommand{ \stautwo}  {\mbox{$\tilde{\tau}_{2}$}}

\newcommand{ \schionezero }{\mbox{$\tilde{\chi}_{1}^{0}$}}
\newcommand{ \schitwozero }{\mbox{$\tilde{\chi}_{2}^{0}$}}
\newcommand{ \schithreezero }{\mbox{$\tilde{\chi}_{3}^{0}$}}

\newcommand{ \schionepm }{\mbox{$\tilde{\chi}_{1}^{\pm}$}}
\newcommand{ \schionemp }{\mbox{$\tilde{\chi}_{1}^{\mp}$}}

\newcommand{ \isajet }    {{\tt ISAJET}}
\def\Journal#1#2#3#4{{#1}{\bf #2} (#4) #3}
\def \PRL      {Phys. Rev. Lett.~}

\def \PRD      {Phys. Rev. D}
\def \PL       {Phys. Lett.~}
\def \PLB      {Phys. Lett. B}
\def \NPB      {Nucl. Phys. B}
\def \PRep     {Phys. Rep.~}

\def \etal     {\relax\ifmmode{et \; al.}\else{$et \; al.$}\fi}

\def \calR     {\relax\ifmmode{{\cal R}}\else{${\cal R}$}\fi}
\begin{document}

\begin{flushright}
MIFP-0606\\
March 16, 2006 \\
\end{flushright}
\vspace*{2cm}

\begin{center}

{\baselineskip 25pt

\large{\bf Detection of SUSY
in the Stau-Neutralino Coannihilation Region
at the LHC} \\
}

\vspace{1cm}

{\large Richard Arnowitt,$^1$ Bhaskar Dutta,$^1$ Teruki Kamon,$^1$
\\ Nikolay Kolev,$^2$ David Toback$^1$} \vspace{.5cm}

{ \it $^1$ Department of Physics, Texas A\&M University, College Station, TX 77843-4242, USA\\
\it $^2$Department of Physics, University of Regina, Regina, SK S4S 0A2, Canada }
\vspace{.5cm}

\vspace{1.5cm} {\bf Abstract}
\end{center}

We study the feasibility of detecting the stau neutralino
($\stauone$-$\schionezero$) coannihilation region at the LHC using tau
($\tau$) leptons. The signal is characterized by multiple  low energy
$\tau$ leptons from $\schitwozero \rightarrow \tau \stauone
\rightarrow \tau\tau \schionezero$ decays, where the $\stauone$ and
$\schionezero$  mass difference ($\dM$)  is constrained to be 5-15 GeV
by current experimental bounds including the bound on the amount of
neutralino cold dark matter. Within the framework of minimal
supergravity models, we show that if hadronically decaying  $\tau$'s
can be identified with~50\% efficiency for visible $\pt >20$ GeV the
observation of such signals is possible in the final state of two
$\tau$ leptons plus large missing energy and two jets. With a gluino
mass of 830 GeV the signal can be observed with as few as 3-10 \invfb\
of data (depending on the size of $\dM$). Using a mass measurement of
the $\tau$ pairs with 10 \invfb\ we can determine $\dM$ with a
statistical uncertainty of~12\% for $\dM$ = 10 GeV and an additional
systematic uncertainty of~14\% if the gluino mass has an uncertainty
of 5\%.

\thispagestyle{empty}

\bigskip
\newpage

\addtocounter{page}{-1}

\section{Introduction}

The advent of supersymmetry (SUSY) has led to the construction of
particle theory models linking a remarkably wide
 range of physical phenomena.
While initially proposed on anesthetic grounds that nature should be
symmetric between fermions and bosons, the fact that supersymmetry
allows for the cancelation of the
 quadratic Higgs divergence allows the building of consistent models
valid up to the  grand unification (GUT) or Planck scales.
 The extension of supersymmetry to a local gauge theory,
supergravity \cite{sugra01,sugra02}, led to the development of
GUT models giving a description of physics from
$M_{\rm GUT}$ down to the electroweak scale and
 incorporating the successes of the Standard Model (SM)~\cite{sugra1,sugra2,nilles}.
Subsequently, LEP data experimentally confirmed the validity of the
idea of SUSY grand unification. Further, an additional feature of SUSY
is that models with
$R$-parity invariance automatically give rise to a candidate,
the lightest neutralino ($\schionezero$),
for the astronomically observed cold dark matter (CDM)
deeply linking particle physics with cosmology, and
detailed theoretical calculations \cite{darkrv} confirm
that GUT models can also achieve the experimentally observed
amount of dark matter \cite{sp} in a natural way.
A large number of
experiments are now under way to try to detect SUSY dark matter in the Milky Way.
Thus it is possible to construct models that encompass the full energy
range of particle physics and simultaneously reach back
into the early universe at times $~10^{-7}$ seconds
after the Big Bang.

If supersymmetry ideas are correct, then the Large Hadron Collider
(LHC) is expected to produce the $\schionezero$ particles. Similarly,
the large one ton dark matter detectors, currently under development,
will cover a large amount of the allowed SUSY parameter space and may
detect directly the Milky Way dark matter particles. Theory predicts
that these two should be the same, and the question arises as to how
one might verify this. The direct approach to this problem would
involve having the dark matter detectors measure the nuclear
differential cross section for the incident dark matter particle
scattering from the detector atomic nuclei, and compare these with the
differential cross sections occurring for the neutralinos produced in
the LHC. Such measurements of course may be many years in the future,
and one would like to see if more immediate measurements might give
strong indications for the equivalence of the astronomical and
accelerator phenomena. To investigate this it is necessary to chose a
SUSY model, and we consider in this paper mSUGRA~\cite{sugra1,sugra2}
(which has been used in many other LHC calculations).

The  allowed mSUGRA parameter space, at present, has three distinct
regions~\cite{darkrv}: (i)~the stau neutralino
($\stauone$-$\schionezero$) coannihilation region where $\schionezero$ is
the lightest SUSY particle (LSP), (ii)~the $\schionezero$ having a
larger Higgsino component (focus point) and (iii)~the scalar Higgs
($A^0$, $H^0$) annihilation funnel
(2$M_{\schionezero}\simeq M_{A^0,H^0}$).
These three regions have been selected out by the CDM
constraint. (There stills exists a bulk region where none of these
above properties is observed, but this region is now very small due to
the existence of other experimental bounds.) The distinction between
the above regions can not be observed in the dark matter direct
detection experiments.  It is therefore important to investigate
whether the dark matter allowed regions can be observed
 at the LHC where the particles will be
 produced directly and their  masses will be measured.
The three dark matter allowed regions need very precise measurements
at the colliders to confirm which is correct.

In this paper we choose to work with the $\stauone$-$\schionezero$
coannihilation region at the LHC.
 We note that many SUGRA
models possess a coannihilation region and if the muon magnetic moment
anomaly, $a_{\mu}$, maintains, it is the only allowed region for
mSUGRA. Coannihilation is characterized by a mass difference ($\dM$)
between $\stauone$ and $\schionezero$ of  about 5-15 GeV in the
allowed region. This narrow mass difference is necessary to allow the
$\stauone$'s to coannihilate in the early universe along with the
$\schionezero$'s in order to produce the current amount of dark matter
density of the universe. Thus if this striking near degeneracy between
$\stauone$ and $\schionezero$ is observed at the LHC, it would be a
strong indirect indication that the $\schionezero$ was the
astronomical dark matter particle. The coannihilation region has a
large extension for the gaugino mass \mhalf, up to  1-1.5 TeV,
 and can be explored at the LHC unless $\tanb$ is very large.

We show here the feasibility of detecting the signal from the
coannihilation region in the first few years of  LHC running. The main
difficulty, however, in probing this region is the small $\dM$ value.
Staus from $\stauone \rightarrow \tau \schionezero$ decays
 generates signals with very low energy tau ($\tau$)
leptons and thus makes it difficult to discover this region at any
collider due to the large size of the background  events from the
SM and SUSY processes. Signals with $\tau$ leptons in the final states
arising from the $\stauone$, $\schitwozero$, and $\schionepm$ decays
have been studied before in the context of the LHC for a large $\Delta
M$ value ($\sim$50 GeV)~\cite{Hinchliffe_Paige_2000}.
A detailed Monte
Carlo (MC) study has been performed to determine SUSY masses by
fitting various mass distributions in SUSY particle
decays~\cite{Heinemann}, but at a pure generator level without any
detector simulation such as effects by reconstruction,
mis-identification, and the SM background. Those studies do not answer
whether we can detect the coannihilation region at the LHC. The
primary feature of this region is that the branching fraction of the
staus and gauginos to $\tau$ leptons is very large, and that
the kinematics of small $\dM$ values produce these $\tau$'s with very
low energy. Both conspire to make any analysis difficult as there can
be very large backgrounds for low energy $\tau$ final states. However,
they do suggest possibilities for measurements. Previously we have
shown the procedure to measure small $\dM$ at an International
Linear Collider (ILC) with an accuracy of about 10\% for our benchmark
points~\cite{ilc_KADKstudy}. But the ILC is still in the proposal
stage. Thus it is important to see if the coannihilation signal can be
observed at the LHC.

We describe here an experimental prescription for detecting SUSY
signal in the coannihilation region and measuring small $\dM$ at the
LHC which provides a necessary condition to be  in the coannihilation
region. We first discuss the available mSUGRA parameter space in
Sec.~2 and the sparticle masses in this region. In Sec.~3 we detail a
search for $\tau$ pairs from $\schitwozero \rightarrow \tau \stauone
\rightarrow \tau\tau \schionezero$ decays and select events in the two
$\tau$ leptons plus large \et\ jet(s) and large missing transverse
energy (\met) final state. In particular, we motivate our selection
criteria, describe the $\tau\tau$  mass and counting observables that
would provide consistency checks and other evidence, and indicate the
amount of luminosity needed to establish a 5$\sigma$ significance. In
this section it will become clear that a primary consideration for
this analysis to have sensitivity is for the LHC detectors to have
efficiency for $\tau$'s with \pt\ $>$~20 GeV. In Sec.~4, we  discuss a
method  of measuring $\dM$ and show the potential accuracy. We
conclude in Sec.~5.

\section{mSUGRA Model in the Coannihilation Region}

The mSUGRA model depends upon four parameters in addition to those of
the SM. They are $m_0$, the universal scalar soft breaking parameter
at $M_{\rm GUT}$; $m_{1/2}$, the universal gaugino mass at $M_{\rm
GUT}$; $A_0$, the universal cubic soft breaking mass at $M_{\rm GUT}$;
and $\tan\beta = \langle \hat{H}_1 \rangle / \langle \hat{H}_2
\rangle$ at the electroweak scale, where $\hat{H}_{1}$ ($\hat{H}_{2}$)
gives rise to up-type (down-type) quark masses. The model parameters
are already significantly constrained by different experimental
results. Most important for limiting the parameter space are: (i)~the
light Higgs mass bound of $M_{h^0} > 114$~GeV from LEP~\cite{higgs1},
(ii)~the $b\rightarrow s \gamma$ branching ratio bound of
$1.8\times10^{-4} < {\cal B}(B \rightarrow X_s \gamma) <
4.5\times10^{-4}$ (we assume here a relatively broad range, since
there are theoretical errors in extracting the branching ratio from
the data)~\cite{bsgamma}, (iii)~the 2$\sigma$ bound on the dark matter
relic density: $0.095 < \Omega_{\rm CDM} h^2 <0.129$~\cite{sp},
(iv)~the bound on the lightest chargino mass of $M_{\schionepm} >$
104~GeV from LEP \cite{aleph} and (v) the muon magnetic moment anomaly
$a_\mu$,  where one gets a 2.7$\sigma$ deviation from the SM from the
experimental result~\cite{BNL,dav,hag}. Assuming the future data
confirms the $a_{\mu}$ anomaly, the combined effects of $g_\mu -2$ and
$M_{\schionepm} >$ 104~GeV then only allows $\mu >0$.

\begin{figure}
\centerline{
\epsfysize=8.0cm \epsfbox{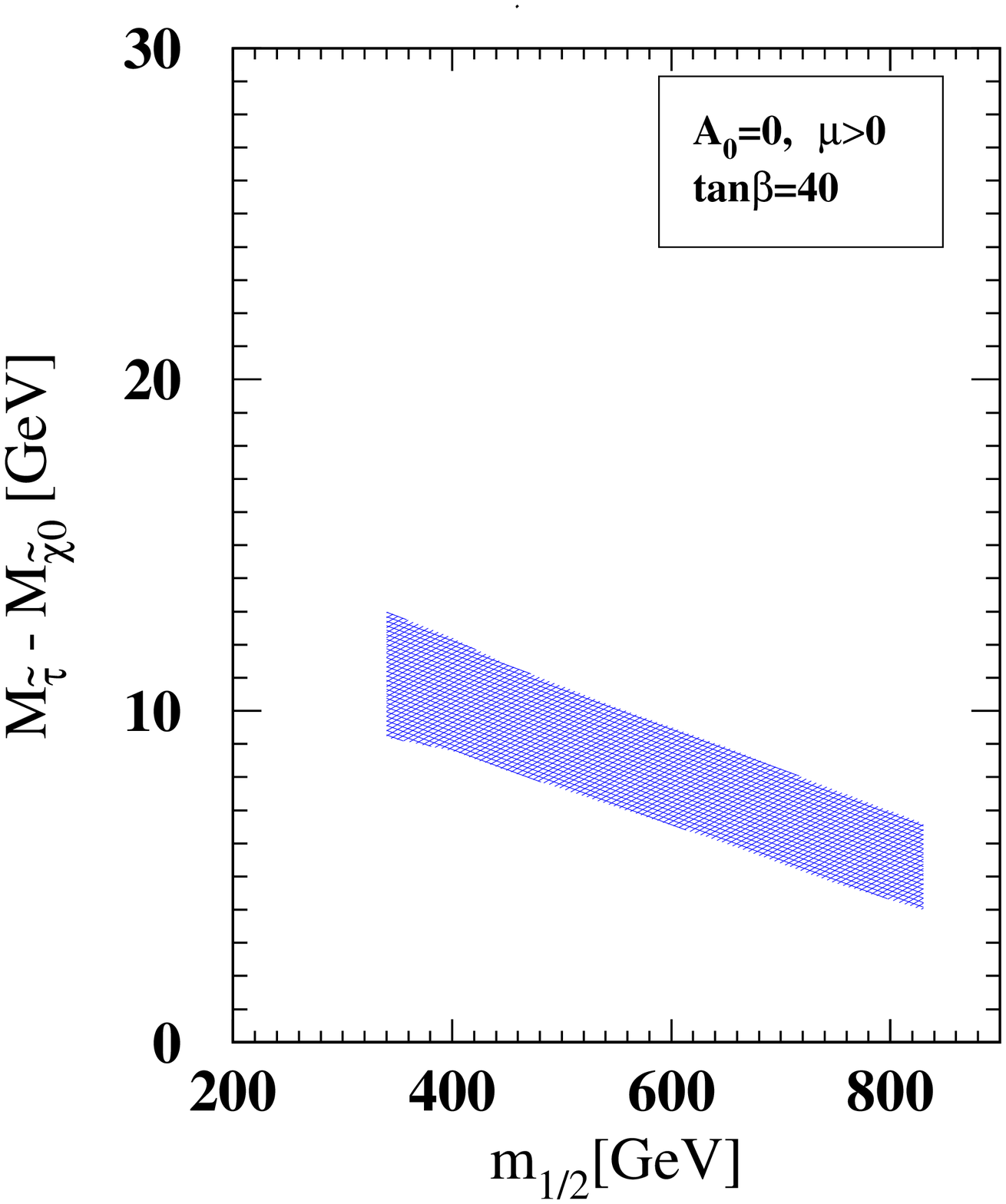} }
\caption{The narrow $\dM$ coannihilation band is plotted as a
function of $m_{1/2}$ for $\tanb$ = 40 with  $A_0  = 0$ and
$\mu >0$. The left end of the band is due to the $b\rightarrow s \gamma$
branching ratio bound  and the right end by $a_\mu<11\times 10^{-10}$.
}
\label{fig:WMAP_allowed_region}
\end{figure}

Figure~\ref{fig:WMAP_allowed_region} shows the range of allowed $\dM$
values in the coannihilation region as a function of $m_{1/2}$ for
$\tanb$ = 40. We see that $\dM$ is narrowly constrained and varies
from 5-15 GeV. Because of the small $\dM$ value, $\tau$'s from
$\stauone \rightarrow \tau \schionezero$ decays are  expected to have
low energy providing the characteristic feature of  the coannihilation
region. From here on, we assume $m_{1/2}=360$ GeV and $\tanb$ = 40 as
a reference point and use it throughout the text unless otherwise
noted. SUSY masses for  five example  those points  are listed in
Table~\ref{tab:SUSYmass} using {\tt ISAJET}
7.63~\protect\cite{isajet}.

\section{Detecting a SUSY Signal in the  Coannihilation Region}

Unlike at the ILC~\cite{ilc_KADKstudy}, it is very difficult to access
direct $\stauonep \stauonem$ production at the LHC because of its
small production cross section relative to the huge QCD backgrounds.
Therefore, we will show  that using copious $\schitwozero$  and decay
via $\schitwozero \rightarrow
    \tau\ \stauone\ \rightarrow\ \tau\ \tau\ \schionezero$
 a significant number of di-tau events can be observed with
an effective  observable.
The end point of the mass distribution is calculated to be
\begin{eqnarray}
\label{eq:mtautaumax}
\mtautaumax  & = &
M_{\schitwozero}
\sqrt{1 - \frac{M_{\stauone}^2}{M_{\schitwozero}^2} }
\sqrt{1 - \frac{M_{\schionezero}^2}{M_{\stauone}^2} }
\end{eqnarray}
which corresponds to the case when the two $\tau$'s are back-to-back
in the $\schitwozero$ rest frame.
To reconstruct the mass as fully as
possible we use hadronically decaying $\tau$'s (\tauh's).
Despite the fact that
the $\tau$'s  lose energy to neutrinos,
the di-tau mass distribution (\mtautauvis) still provides
a peak position (\mtautaupeak)
 that can be directly measured.
The peak position is  below \mtautaumax\ and \mtautauvis\ is
effectively smeared. We next describe how to use these experimental
features to detect the signal.

\begin{table}
\caption{Masses (in GeV) of SUSY particles in five representative
scenarios  for
$m_{1/2}$ = 360~GeV, $\tanb = 40$, $\mu > 0$, and $A_0 = 0$.
These points satisfy all the existing experimental bounds on mSUGRA.
$\mtautaumax$ is the end point of the di-tau mass
distribution from the $\schitwozero$ decays. The production cross
section is nearly same for the five points and is 8.3 pb
for $\mzero$ = 215 GeV.}
 \label{tab:SUSYmass}
\begin{center}
\begin{tabular}{c c c c c c }
\hline \hline
$m_{0}$     &210    &212    &215    &217    &220    \\
\hline
\gluino     &831  &831  &831  &831  &832  \\
\usquarkL   &764  &764  &765  &765  &766  \\
\usquarkR   &740  &740  &741  &741  &742  \\
\stoptwo    &744  &744  &744  &745  &745  \\
\stopone    &578  &578  &579  &579  &580  \\
\stautwo    &331  &332  &333  &334  &336  \\
\seleL      &323  &324  &326  &328  &330  \\
\schitwozero    &266  &266  &266  &266  &266  \\
\seleR      &252  &254  &256  &258  &260  \\
\stauone    &149.9  &151.8  &154.8  &156.7  &159.5  \\
\schionezero    &144.2  &144.2  &144.2  &144.2  &144.2  \\
$\dM (\equiv M_{\stauone} - M_{\schionezero})$   & 5.7    & 7.6  & 10.6  & 12.5  &15.4   \\
\mtautaumax  & 60.0 & 68.3  & 78.7 & 84.1 & 91.2
\\\hline \hline
\end{tabular}
\end{center}
\end{table}

The primary SUSY production processes  at the LHC are $p p \rightarrow
\squark \gluino , \squark \squark , \gluino \gluino$. In each case the
decays proceed via $\squark  \rightarrow  q^{\prime} \schionepm$ or $q
\schitwozero$ (or $\squarkR \rightarrow q \schionezero$); $\gluino
\rightarrow q \bar{q}^{\prime} \schionepm$ or $q \bar{q}
\schitwozero$; and $\gluino \rightarrow \tbar \stopone$ or $\bbar
\sbottomone$ and their charge conjugate states, generally producing
high \et\ jets and gaugino pairs. We are most interested in events
from $\schionezero \schitwozero$, $\schionepm \schitwozero$, or
$\schitwozero \schitwozero$ pairs, where the $\schionezero$ in the
first case is directly from the $\squarkR$ decay. The branching ratio
of $\schitwozero  \rightarrow \tau\ \stauone$ is about 97\% for our
parameter space and is dominant even for large $\mhalf$ in the entire
coannihilation region; the same is true for the $\schionepm
\rightarrow \nu\ \stauone$ decay mode. (It should be noted that both
\seleR\ and \smuR\ are lighter than $\schitwozero$ by about 10~GeV .
However, the branching ratio for $\schitwozero \to e \seleR$ or $\mu
\smuR$ is much less than 1\%.) Since the stau decays via $\stauone
\rightarrow \tau \schionezero$,  we expect inclusive $\schitwozero$
events to include at least two $\tau$ leptons plus large \et\ jet(s)
and large \met\ (from the $\schionezero$).

We consider two experimental scenarios. The first uses the $\met$ +
$\geq$2~jet final state to reduce backgrounds and searches for  the
2$\tau$'s that arises from the decays of \schitwozero.
In each
candidate event,
 all di-tau pairs can then be  searched for a mass peak as evidence of
the \schitwozero\ decay chain. The second option studies gaugino pairs
($\schitwozero \schitwozero$, $\schionepm\schitwozero$) and requires
3$\tau$'s to reduce backgrounds and only 1~jet + $\met$ in the  final
state to regain acceptance. Both final states will be triggered by
requiring large \et\ jet(s) and large \met\ and  such a trigger will
be available at both the ATLAS and the CMS experiments. We focus on
the final event selection and report here on an analysis of the
$\met$ + $\geq$2 jet + $\geq 2\tau$ events
using the \isajet\ event generator
and the ATLAS detector simulation {\tt ATLFAST}~\cite{atlfast}.
The results of the 3$\tau$ analysis will be considered
elsewhere~\cite{3tau_paper}.

To establish a signal in the coannihilation region we require (at
least) (1)~a 5$\sigma$ excess in the counting of \met\ + 2 jet +
$2\tau$  events with \mtautauvis\   below \mtautaumax\ and (2)~the
mass peak to be  consistent with the expectations from $\schitwozero$
decays. The question then becomes  whether such measurements are
possible at the LHC. We begin by addressing the primary experimental
issues: (a) the \pt\ spectrum of $\tau$'s from $\stauone \rightarrow
\tau \schionezero$ which is  expected to be soft and (b) the ability
to select the correct di-tau pairs.

\begin{figure}
\centerline{
\epsfysize=8cm
\epsfbox{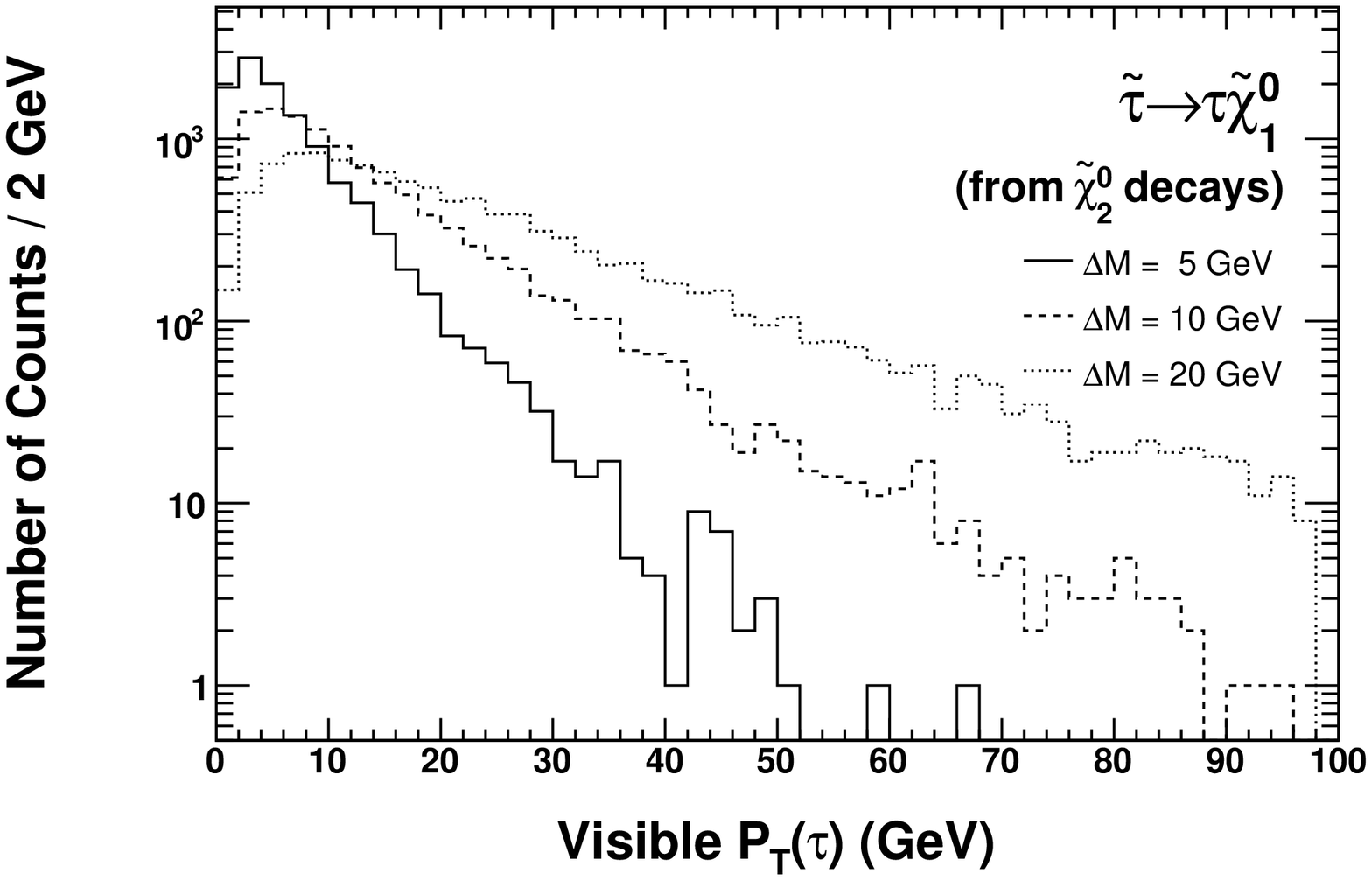} }
\caption{The visible
\pt\ distributions for  hadronically decaying $\tau$ leptons from
$\stauone \to \tau \schionezero$ where the  $\tilde\tau_1$'s are
required to be decays of $\schitwozero \to \tau \stauone$. The curves
show the results for $\dM$ = 5, 10, and 20 GeV.}
\label{fig:visTauEt_mSUGRA}
\end{figure}

We first examine the visible \pt\ (\ptvis) distribution of
$\tauh$'s from  $\stauone \rightarrow \tau \schionezero$ in
$\schitwozero$ decays with $\dM$ = 5, 10, and 20 GeV using \isajet. As
seen in Fig.~\ref{fig:visTauEt_mSUGRA}, even with a small mass
difference, the  $\tau$ is boosted in the cascade decay of the heavy
squark and gluino making it potentially viable. One can already begin
to see the importance of reconstructing the $\tau$'s with
 $\ptvis\ \gtsim\ 20\ \gev$.
From here on, we assume that both the ATLAS and CMS detectors can
reconstruct and identify $\tauh$'s with  $\ptvis > 20\ \gev$. Assuming
the taus are identifiable, we  require two jets with $\et >100\ \gev$
and $\met > 180\ \gev$. These cuts should satisfy the \met\ + jets
trigger at ATLAS and CMS with high enough efficiency that we further
neglect trigger efficiency and bias effects.

\begin{figure}
\centerline{ \epsfysize=5.5cm \epsfbox{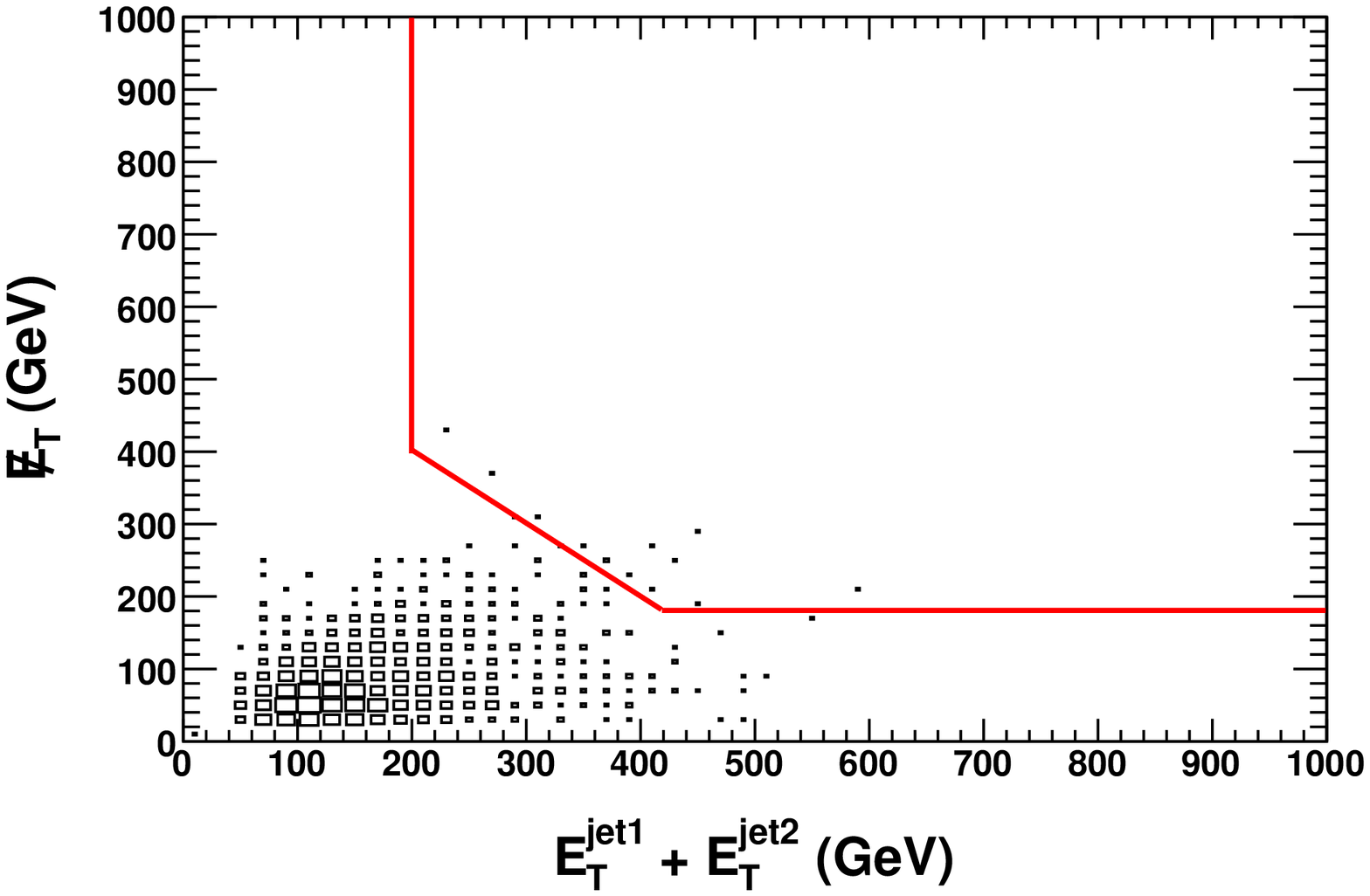}
\epsfysize=5.5cm \epsfbox{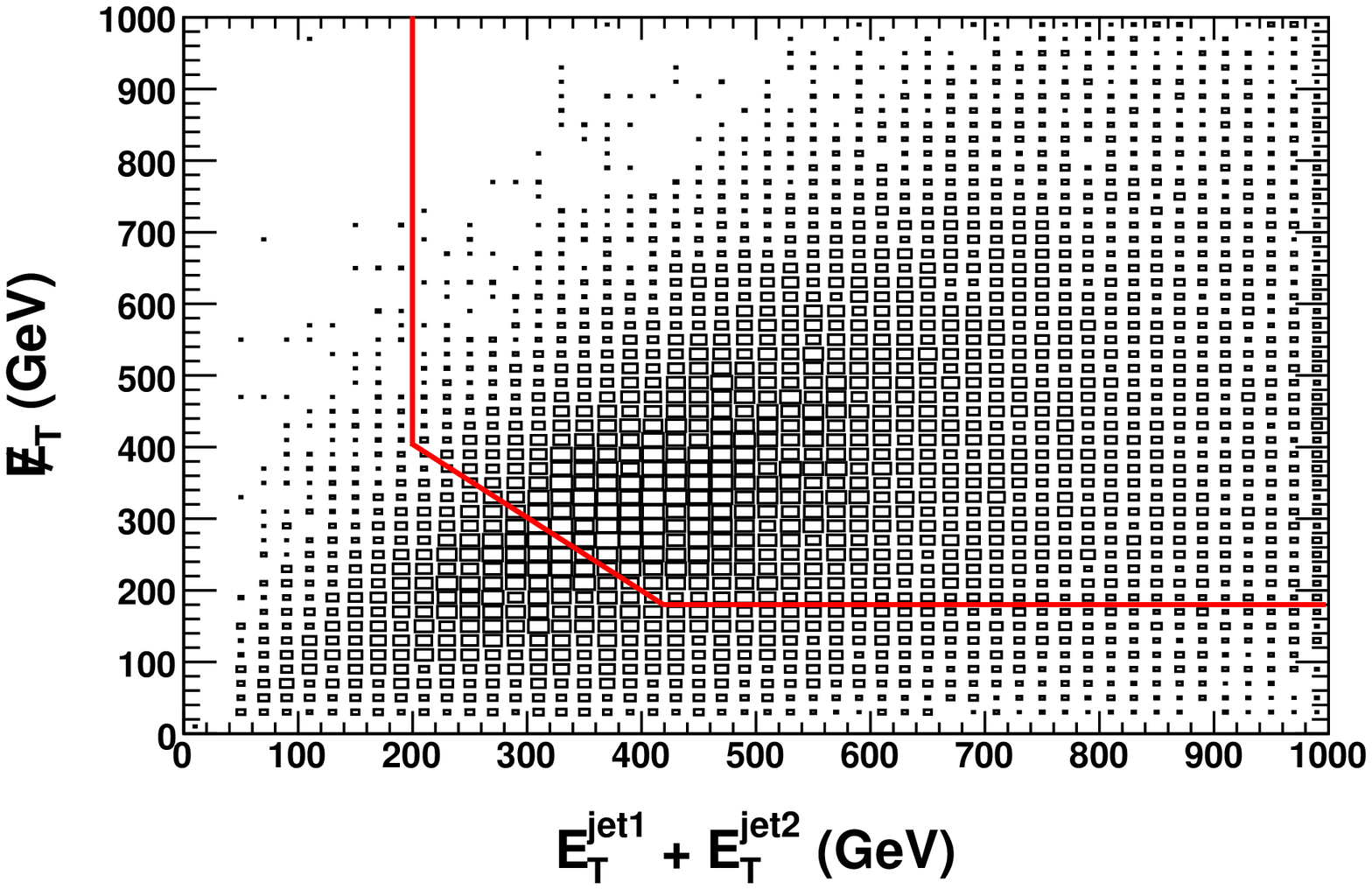}} \caption{Scatter
plots of \met\ vs $E_{\rm T}^{~{\rm jet 1}} + E_{\rm T}^{~{\rm jet
2}}$ for $t\bar t$ (left) and SUSY (right) events after requiring
$2\tau$'s passing all ID and kinematic cuts in the final state. The
SUSY events are for our reference point at $M_{\gluino}$ = 831~GeV and
$\dM = 10.6\ \gev$. We require $E_{\rm T}^{~{\rm jet 1}} > 100\ \gev$,
$E_{\rm T}^{~{\rm jet 2}} > 100\ \gev$, $\met > 180\ \gev$, and
$E_{\rm T}^{~{\rm jet 1}} + E_{\rm T}^{~{\rm jet 2}} + \met
>$ 600 GeV to eliminate the top background.
}
\label{fig:top_vs_SUSY}
\end{figure}

Events are further selected by requiring at least two identified
$\tauh$'s with at least one $\tau$ lepton with $\ptvis > 40\ \gev$
(from $\schitwozero \rightarrow \tau \stauone$) and one with $\ptvis >
20\ \gev$ (from $\stauone \rightarrow \tau \schionezero$).  At this
point, the dominant background is expected to be from  $t\bar t$ pair
production, where each $t$ decays as $t \rightarrow b \tau \nu$ and
produces a final state of $\met + 2 b + 2 \tau$. A correlation plot
between \met\ and $E_{\rm T}^{~{\rm jet 1}} + E_{\rm T}^{~{\rm jet
2}}$ is shown in Fig.~\ref{fig:top_vs_SUSY} for $\ttbar$ and SUSY
events.
  We choose
$E_{\rm T}^{~{\rm jet 1}} + E_{\rm T}^{~{\rm jet 2}} + \met > 600\ \gev$ to further reduce the background.

\begin{figure}
\centerline{ \epsfysize=7cm
\epsfbox{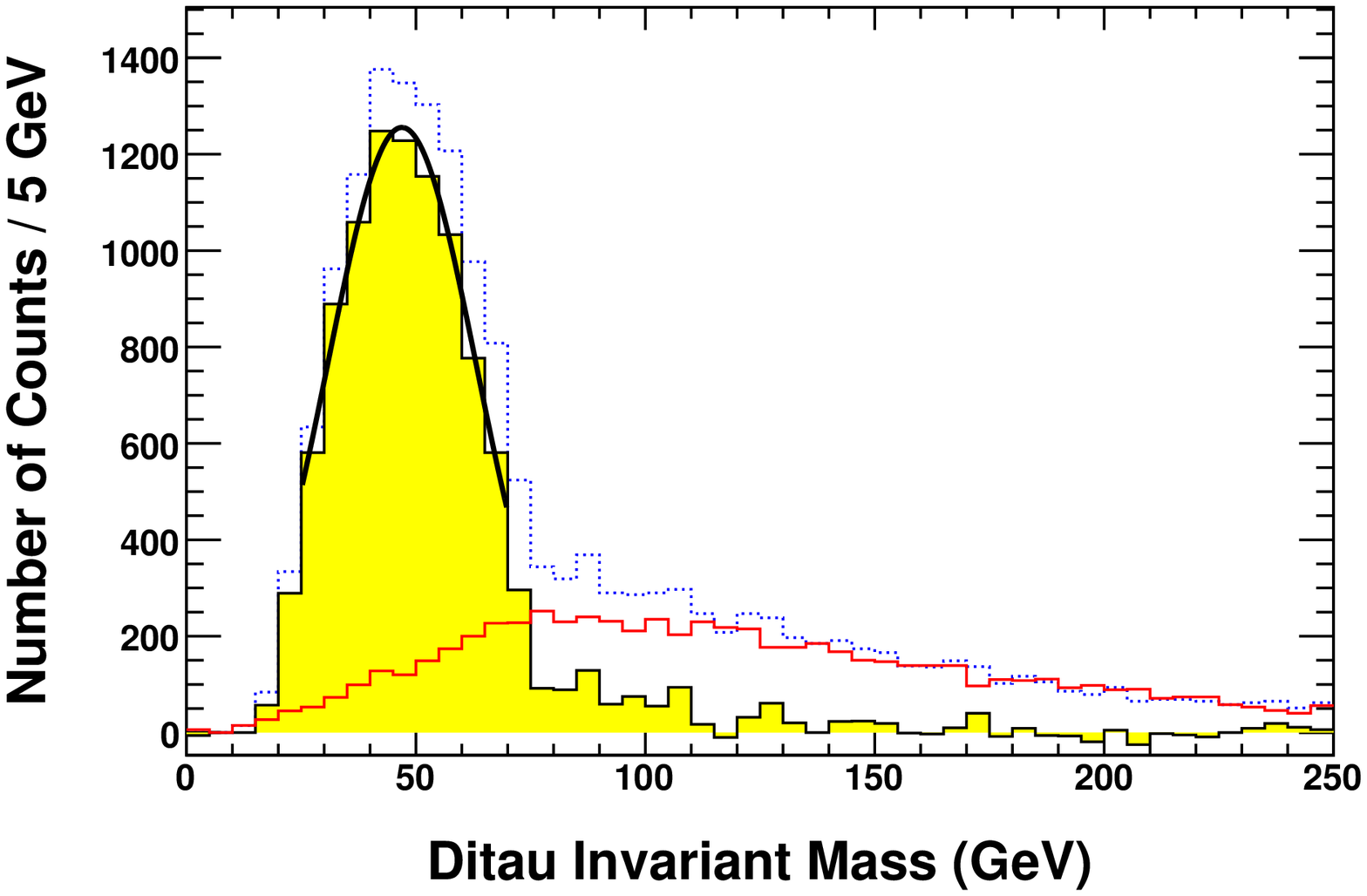} } \caption{The visible
di-tau invariant mass (\mtautauvis) distributions for all di-tau
combinations in inclusive  \met\ + 2j + 2$\tau$ events in the
coannihilation region ($\dM = 10.6\ \gev$; see
Table~\ref{tab:SUSYmass}) with $\epsilon = 100\%$ and \faketau\ = 0\%.
The $\tau$'s are selected with $\ptvis > 20\ \gev$ but requiring at
least one $\tau$ lepton to have $\ptvis > 40\ \gev$. The  endpoint is
consistent with the theoretical value at \mtautaumax = 78.7~GeV. The
dashed and solid open histograms are for OS and LS pairs,
respectively. The gray (or yellow) histogram is for OS$-$LS pairs and
is well fitted to a Gaussian (solid line). The small excess between 80
and 150 GeV is mainly due to $\schithreezero\ {\rm (456\ GeV)}
\rightarrow \tau \stauone, Z \schitwozero, W^{\pm} \schionemp$ decays.
 }
\label{fig:ditau_mass_shape}
\end{figure}

For each candidate  \mtautauvis\ is calculated for every pair of
$\tau$'s in the event  and categorized as opposite sign (OS) or like
sign (LS). The mass distribution for LS pairs is subtracted from the
distribution for OS pairs to extract $\schitwozero$ decays on a
statistical basis. The result is shown in
Fig.~\ref{fig:ditau_mass_shape} where we have assumed the
identification (ID) efficiency ($\epsilon$) to be 100\% and a
probability that a jet is misidentified as $\tauh$ (\faketau) to be
0\%. We see that the non-$\tilde\chi^0_2$  OS  pairs are nicely
canceled with the wrong LS combination pairs and that the OS$-$LS
distribution is well fit to a Gaussian. We note that while the
expected maximum di-tau mass, $\mtautaumax = 78.7\ \gev$, is
consistent with  Fig.~\ref{fig:ditau_mass_shape} and easily
determined, we believe  a full detector simulation is required to show
this in practice. We note for completeness  that there is  a small
second  excess between 80 and 150~GeV, which   is mainly due to
$\schithreezero\ {\rm (456\ GeV)} \rightarrow \tau \stauone, Z
\schitwozero, W^{\pm} \schionemp$ decays.

\begin{figure}
\centerline{ \epsfysize=5.5cm \epsfbox{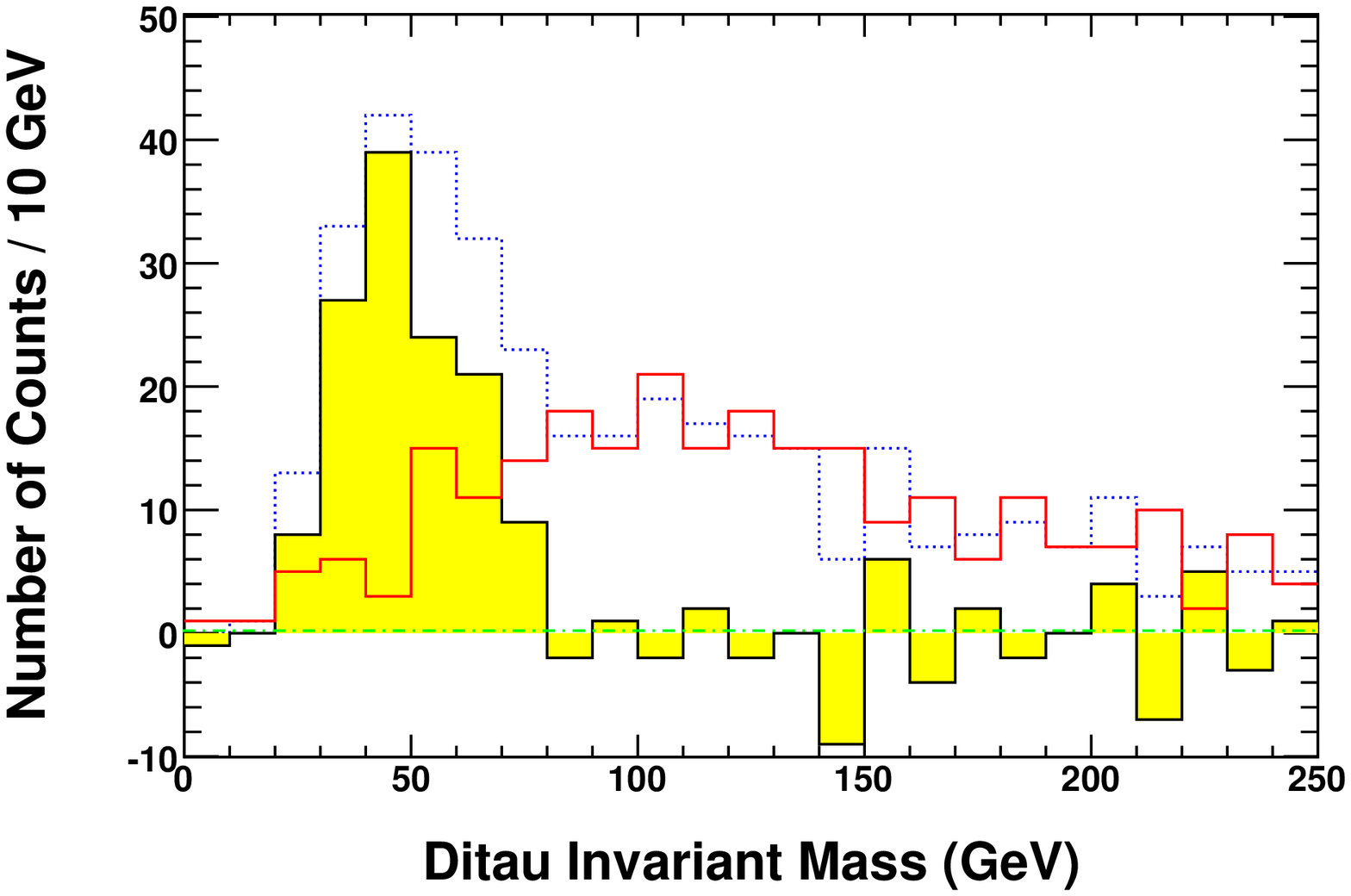}
\epsfysize=5.5cm \epsfbox{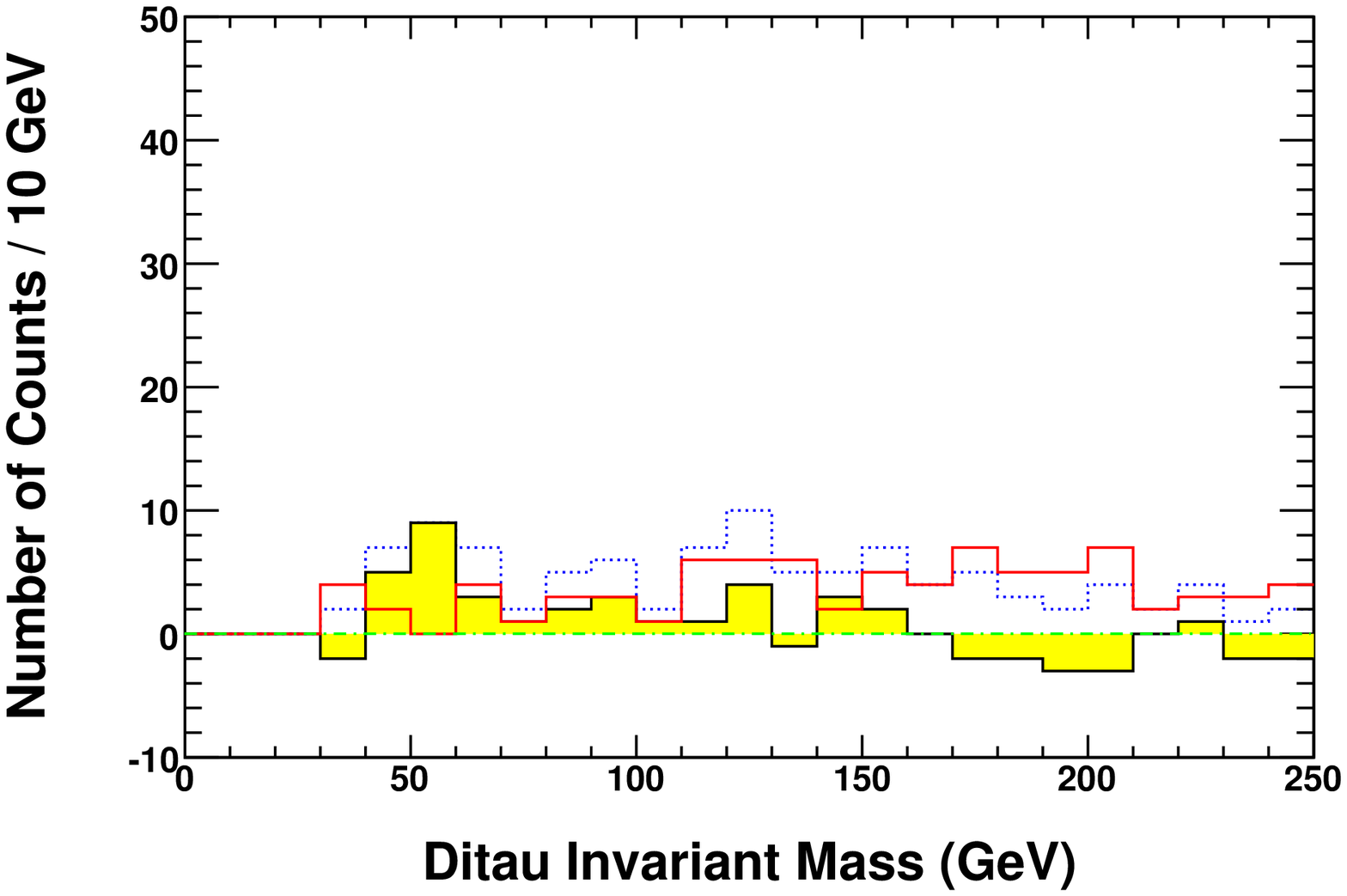} }
\caption{[left] Same as in Fig.~\ref{fig:ditau_mass_shape}, but using
a randomly-selected 10 \invfb\ SUSY sample that is analyzed with
$\epsilon = 50\%$ and \faketau\ = 1\%. \mtautaumax\ = 78.7 GeV can
still be inferred and a clear peak is visible. Note that the $\ttbar$
contribution in $\mtautauvis < 100\ \gev$ is estimated to be a few
OS$-$LS counts and has not been plotted. [right] The same sample but
requiring all $\tau$ candidates to have  $\ptvis > 40\ \gev$.  The
characteristic di-tau mass distribution has disappeared.
 }
\label{fig:mass_ditau}
\end{figure}

We next assume $\epsilon$ = 50\% and \faketau\ = 1\%, based on the CDF
measurement~\cite{cdfprl_himass_ditau}, and show that these results
are not significantly affected by these experimental conditions except
to quadruple the needed luminosity to establish a signal.
Figure~\ref{fig:mass_ditau} shows the \mtautauvis\ distributions for
OS, LS, and OS$-$LS pairs for a hypothetical  10 \invfb\ sample to
allow visual confirmation of the peak taking into account statistical
fluctuations. Since the number of $\ttbar$ events that survive our
selection cuts is expected to be a few OS$-$LS counts per 10 \invfb\
we do not consider them further. The excess between 80 and 150 GeV
seen in Fig.\ref{fig:ditau_mass_shape} becomes statistically
insignificant. It should be noted that a set of similar kinematical
cuts is also used in Ref.~\cite{Hinchliffe_Paige_2000} and the SM
background was also found to be small. To further illustrate the
importance of reconstructing $\tau$'s with $\ptvis \gtsim\ 20\ \gev$,
the RHS of Fig.~\ref{fig:mass_ditau} shows that the  mass peak
disappears for the same sample, if we require all $\tau$'s to be above
40 GeV.

\begin{figure}
\centerline{
\epsfysize=7cm
\epsfbox{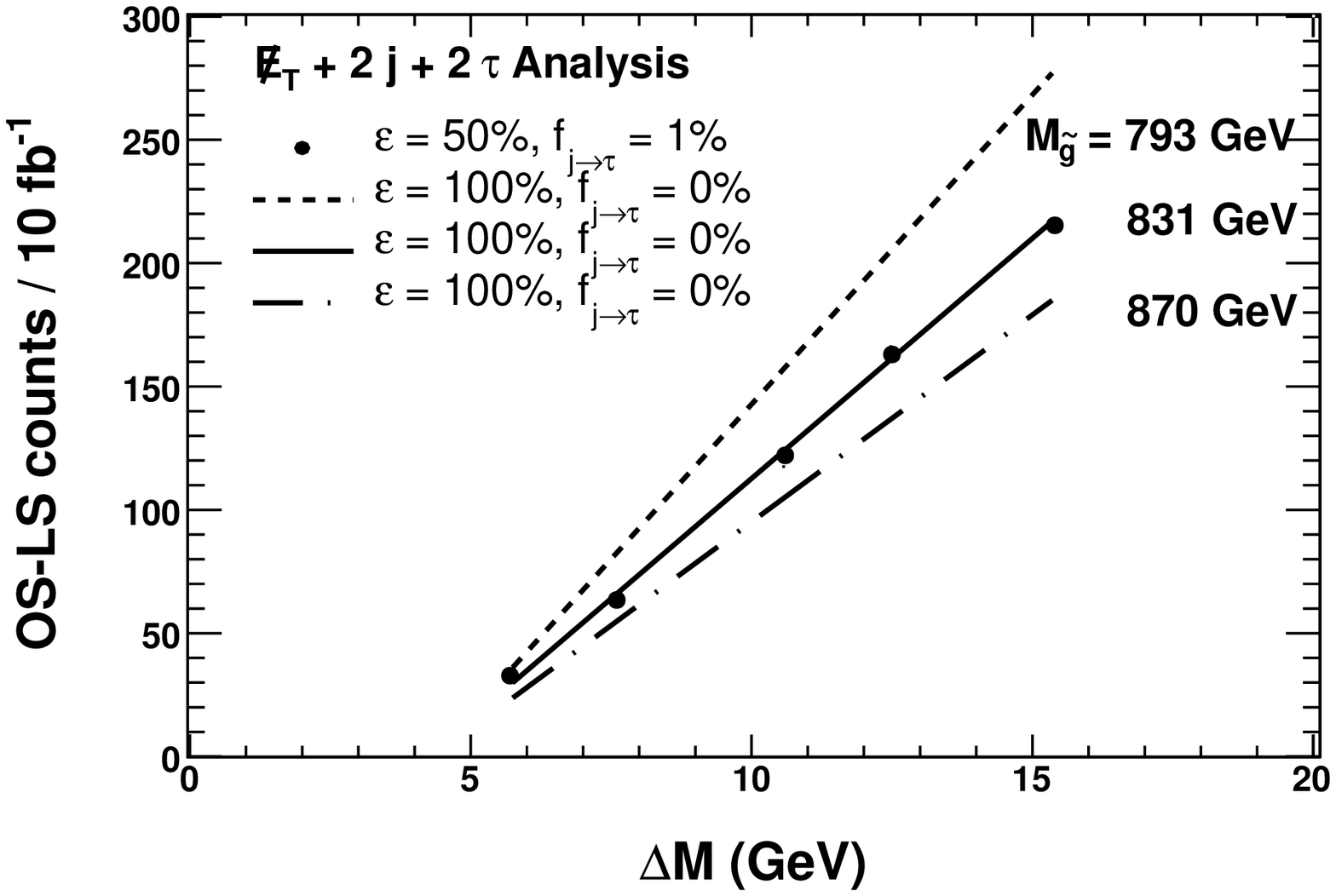}
}
\caption{The mean number of OS$-$LS counts below the end point
per 10 \invfb\ of data
as a function of $\dM$.
The solid circles are for $\epsilon = 50\%$ and \faketau\ = 1\%.
The solid line
is  the case
with $\epsilon = 100\%$ and \faketau\ = 0\%,
 but with the number of OS$-$LS counts  scaled down by a factor of 4
due to $1/\epsilon^2$.
The uncertainty of each circle is about its own size
based on 200 \invfb\ SUSY MC samples.
The dashed and dash-dotted lines are when the gluino masses are varied
by about 5\%.
 }
\label{fig:N_vs_dM}
\end{figure}

For any  coannihilation signal to be established a necessary
(but not sufficient) condition is that
an excess of signal events above
background should be readily observed.
Figure~\ref{fig:N_vs_dM} shows
the expected yield of OS$-$LS counts below the end point per 10
\invfb\ as a function of $\dM$.
For comparison, we show  the results for
 $\epsilon = 50\%$ and
\faketau\ = 1\%, along with the  ideal $\tau$ ID condition of
$\epsilon = 100\%$ and \faketau\ = 0\% but scaled down by a factor 4
due to $1 / \epsilon^2$.  It is remarkable to see that there is no
significant difference between two cases, suggesting that the counting
is very insensitive to the fake rate. This suggests that the counting
nearly perfectly reflects the counting of $\schitwozero$'s if one
takes into account the $\tau$ ID efficiency. This also indicates that
in the signal region there is only a small contamination from the
wrong combination events. Figure~\ref{fig:N_vs_dM} also shows the
numbers of OS$-$LS counts if the gluino mass is varied by
approximately $\pm 5\%$
 (and thus also the $\schitwozero$ and $\schionezero$ masses), but
 maintaining the same $\dM$ values.
 Heavier (lighter)
gluino mass is reflected as a smaller (larger) production cross
section, thus with a smaller (larger) yield.

\begin{figure}
\centerline{ \epsfysize=9cm
\epsfbox{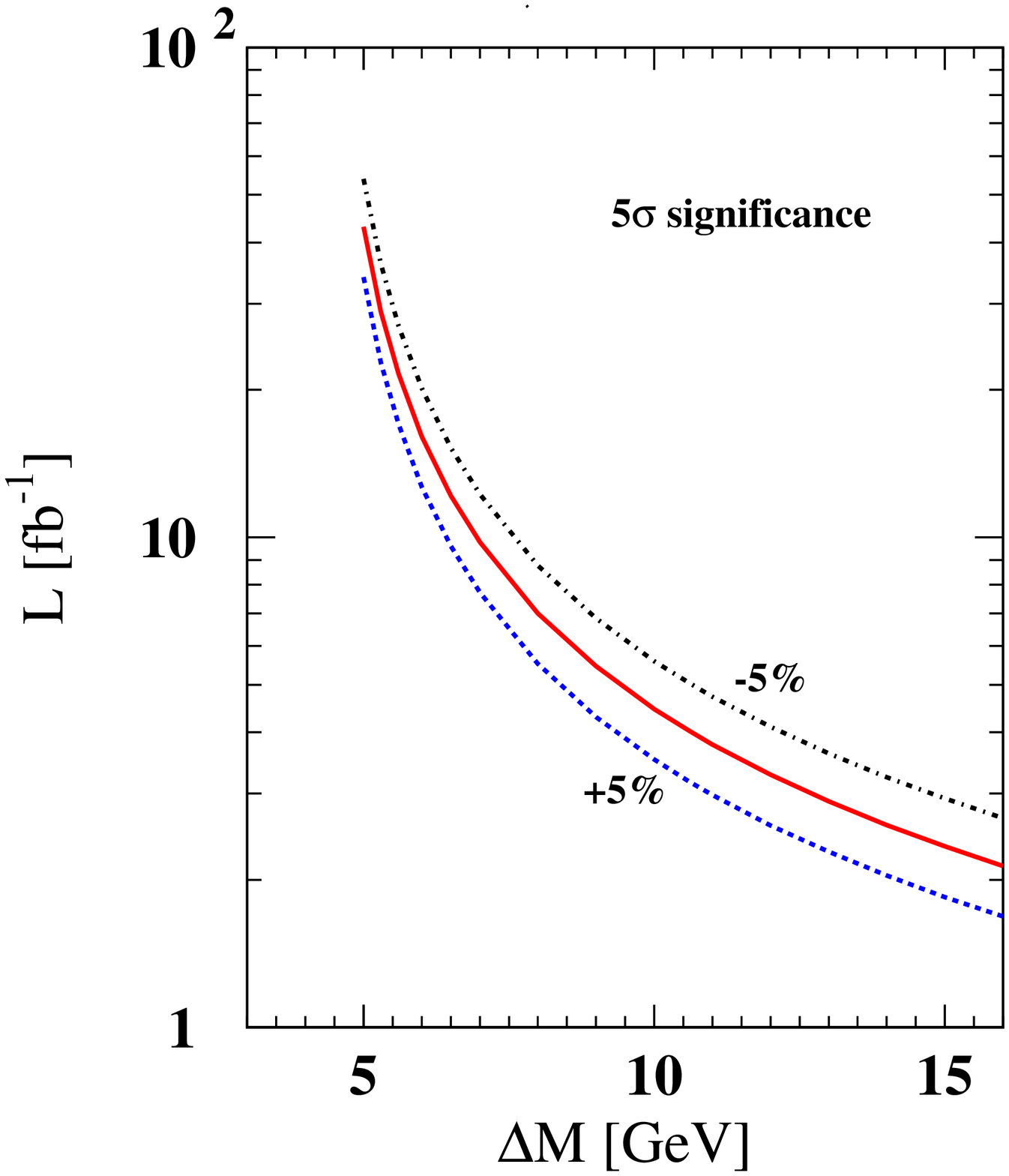} } \caption{The
luminosity necessary to establish a 5$\sigma$ significance in the
number of  OS$-$LS counts for the coannihilation region. The band
reflects a variation  due to the gluino mass by $\pm5\%$ from a
nominal gluino mass of 831 GeV in our reference model (see
Table~\ref{tab:SUSYmass}). } \label{fig:Lum25}
\end{figure}

We next determine
the luminosity necessary to
observe OS$-$LS counts with a 5$\sigma$ significance. We define
$\sigma \equiv (N_{\rm OS} - N_{\rm LS})/\sqrt{N_{\rm OS} + N_{\rm LS}}$,
again require  $\mtautauvis < \mtautaumax$, and use
the results in Fig.~\ref{fig:N_vs_dM} to parameterize
the number of OS$-$LS counts
 as a function of $\dM$~\cite{LHCstau_parameterization}.
The luminosity required for as  a 5$\sigma$ excess as a function
  $\dM$ is shown in Fig~\ref{fig:Lum25}.
 We conclude that the characteristic coannhilation
signal ($\dM$ = 5-15 GeV) can be detected at the LHC with as few
as 3 \invfb\ if $\dM$ is large ($\simeq$15 GeV) and with 10 \invfb\ if
$\dM$ is small (assuming  the SUSY GUT models are correct and the
scale is $M_{\gluino} = 831\ \gev$).

\begin{figure}
\centerline{ \epsfysize=8cm \epsfbox{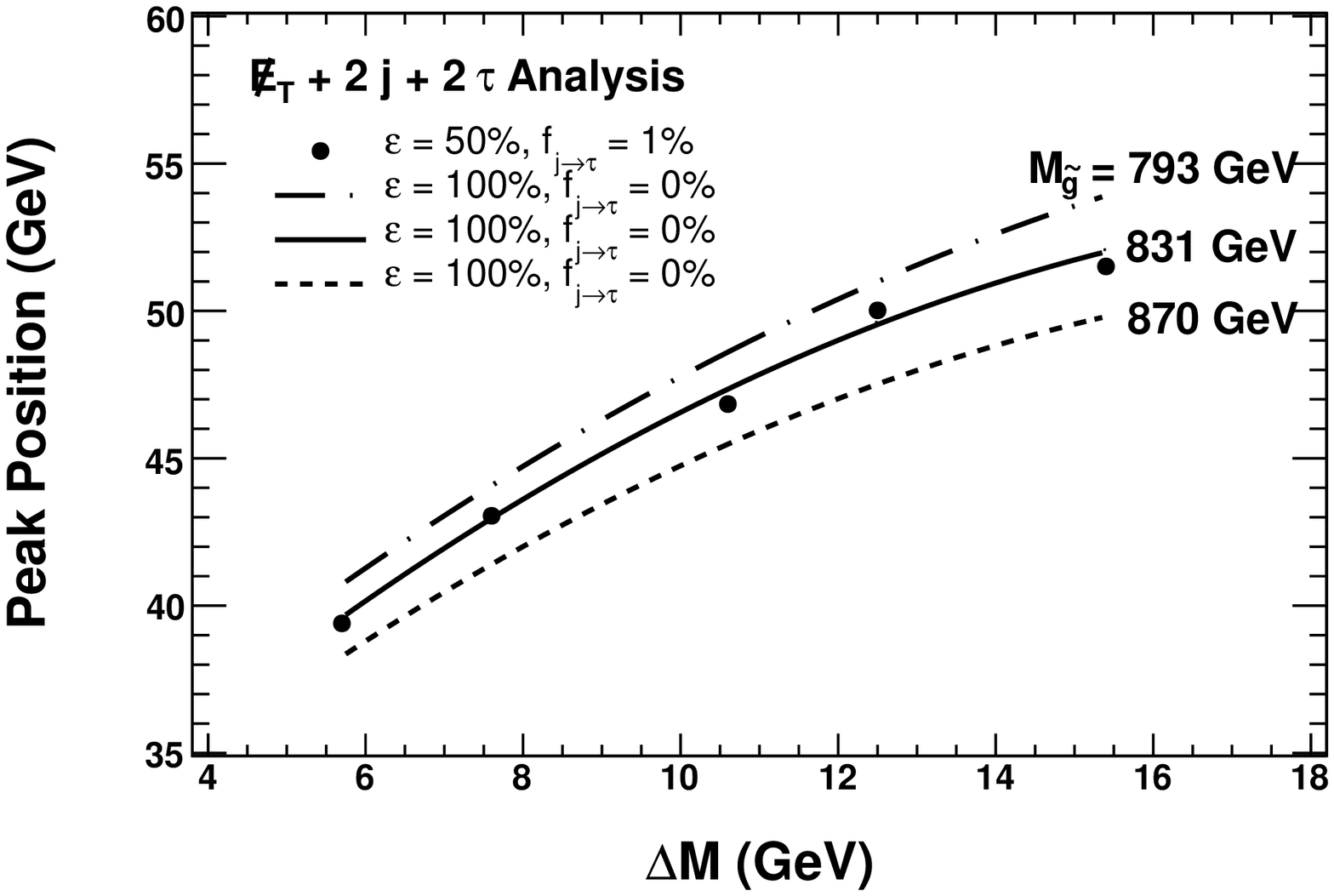} }
\caption{The peak  of the \mtautauvis\ distribution as a function of
$\dM$. The solid circles are for $\epsilon = 50\%$ and \faketau\ =
1\%, compared to the solid line for the case with $\epsilon = 100\%$
and \faketau\ = 0\%. The statistical uncertainty on each circle is
about the size of the circle from 200 \invfb\ SUSY MC samples. The
dashed and dash-dotted lines are when the $\gluino$ masses are varied
by about 5\% and can be understood as coming from the change in the
$\schitwozero$ mass for a fixed $\dM$, since mSUGRA provides  a fixed
relationship between $M_{\gluino}$ and $M_{\schitwozero}$ (See Eq.1).
Note that the variation due to the fake rate  is negligible.}
\label{fig:mass_dM}
\end{figure}

The second criteria we consider to help establish the coannihilation
region signal is the observation of \mtautaupeak\ at the correct
location. Within mSUGRA models, since the $\gluino$, $\schionezero$
and $\schitwozero$ masses are related,
\mtautaumax\ changes with $M_{\stauone}$
for a given $M_{\schionezero}$. This also
means \mtautaupeak\ changes with $\dM$.
Figure~\ref{fig:mass_dM} shows
\mtautaupeak\  as a function of $\dM$;
any experimental observation would need to be consistent with this
small range of 39-52 GeV for the case of $M_{\gluino} = 831\ \gev$.

We next comment on two effects that could that could change the
expected \mtautaupeak\ value. The first is that
since there is a direct relationship between $M_{\schitwozero}$ and
$M_{\gluino}$, from Eq.~\ref{eq:mtautaumax},
the peak position is expected to be a
function of $M_{\gluino}$ for a given $\dM$.
If $M_{\gluino}$ is
changed from our nominal value but the $\dM$ value is maintained, we
find
 $\delta_{\mtautaupeak}  / \mtautaupeak  \simeq
 \delta_{\mtautaumax} / \mtautaumax
= 0.718\ \delta_{M_{\gluino}} / M_{\gluino}$.
This is directly observed in Fig.~\ref{fig:mass_dM}
and effectively expands the
observation range to 38-54 GeV. We also study the potential
impact of varying $\tau$ ID conditions. On the same figure we show
(a)~$\epsilon = 100\%$ and \faketau\ = 0\% and  (b)~$\epsilon = 50\%$
and \faketau\ = 1\% and see that there is no systematic peak shift due
to the jet $\rightarrow \tau$ fake rate. This can be explained by the
fact that the jet misidentified as a $\tau$ has no correlation with
the $\tau$ leptons from the $\schitwozero$ decay; both OS and LS
di-tau mass distributions are affected equally in its shape, and thus
will cancel in the difference to  good approximation.

As an aside we note that there is a non-coannihilation region where a
low energy $\tau$ lepton is expected when the $\stauone$ mass  is
close to the $\schitwozero$ mass instead of the $\schionezero$. In
this ``inverted mass'' scenario, $\dM_{\rm inv} \equiv
M_{\schitwozero} - M_{\stauone}$ = 5-15 GeV. SUSY masses in two
particular cases are listed in Table~\ref{tab:SUSYmass_Inv}, compared
to our reference point ($\dM$ = 10.6 GeV and \mtautaumax\ = 78.7 GeV).
The di-tau mass distribution for the events would look similar to
those  in the coannihilation region, but we expect a difference in the
invariant mass of the jet-\tauh-\tauh\ system in the $\squark
\rightarrow q \schitwozero$ decay where the squarks are heavier by
about 30~GeV compared to those of the reference point.  A full study
is beyond a scope of this study, and we will address this in a future
paper. We note, however, that the above two cases do not satisfy the
relic density constraint and hence are inconsistent with cosmology.
Also, while the small $\dM$ of the coannihilation region is natural
for mSUGRA to be consistent with WMAP data, the inverted mass scenario
further requires an artificial fine tuning to make $M_{\schitwozero} -
M_{\stauone}$ small.

Having shown that we can establish a  5$\sigma$ excess and that  the
peak and end point of di-tau mass distribution are consistent with
expectations and only occupy a small region of possible values, we
next discuss the prospects for measuring
 $\dM$  with 10 \invfb\ of data using the peak position of the di-tau mass distribution.

\section{Measuring $\mathbf{\Delta {\it M}}$ Using 10 fb$\mathbf{^{-1}}$}

The  \met + 2j + 2$\tau$ data set  provides an opportunity to make a
$\dM$ measurement. Since \mtautaupeak\ varies with $\dM$, a
measurement of \mtautaupeak\ combined with a measurement of
$M_{\gluino}$ performed elsewhere  allows for a measurement of $\dM$.
It should be noted that from Fig.~\ref{fig:mass_dM}, the peak position
varies with $M_{\gluino}$. This arises from the mSUGRA relations
$M_{\gluino} \simeq 3.1~M_{\schitwozero}$ and the fact that the
gaugino masses enter in the endpoint relation of
Eq.~\ref{eq:mtautaumax}. We note for completeness that a similar
measurement could be done with the results of Fig.~\ref{fig:N_vs_dM}.
This will be done is a separate paper~\cite{3tau_paper}.

The statistical uncertainty of the $\dM$ measurement is dominated by
the precision of the $M_{\tau\tau}^{\rm peak}$ measurement. The
statistical uncertainty in the measurement of  \mtautaupeak\ is
typically in the 1 GeV range for 10 \invfb. Numerically it is given by
$\sigma_{M}/\sqrt{N}$, where $\sigma_{M}$ is the r.m.s. of the mass
distribution (see Fig.~\ref{fig:ditau_mass_shape}) and is of the order
of 10 GeV, and $N$ is the number of OS$-$LS counts in the peak region
for a given luminosity (see Fig.~\ref{fig:N_vs_dM}). For a more
detailed parameterizations see Ref.~\cite{LHCstau_parameterization}.
The systematic uncertainty due to the uncertainty in the assumed value
of $M_{\gluino}$ which changes the \mtautaupeak\ vs $\dM$ relationship
(see Fig.~\ref{fig:mass_dM}) is also of the order of 1 GeV for  a 5\%
uncertainty in the gluino mass. The uncertainty due to \faketau\ does
not affect the peak measurement.

\begin{table}
\caption{Masses (in GeV) of SUSY particles in two ``inverted mass"
scenarios  for $m_{1/2}$ = 360~GeV, $\tanb = 40$, $\mu > 0$, and $A_0 = 0$.}
\label{tab:SUSYmass_Inv}
\begin{center}
\begin{tabular}{c c c}
\hline \hline
$m_{0}$     &323.7    &315.2       \\
\hline
\usquarkL   &798   &795  \\
\stopone    &598   &560  \\
\schitwozero&266.0 &266.0 \\
\stauone    &255.4 &247.8 \\
\schionezero&144.2  &144.2   \\
$\dM_{\rm inv} (\equiv M_{\schitwozero} - M_{\stauone})$   & 10.6   & 18.2  \\
$M_{\tau\tau}^{\rm max}$ & 61.4 & 78.7\\
\hline \hline
\end{tabular}
\end{center}
\end{table}

\begin{figure}
\centerline{
\epsfysize=10cm
\epsfbox{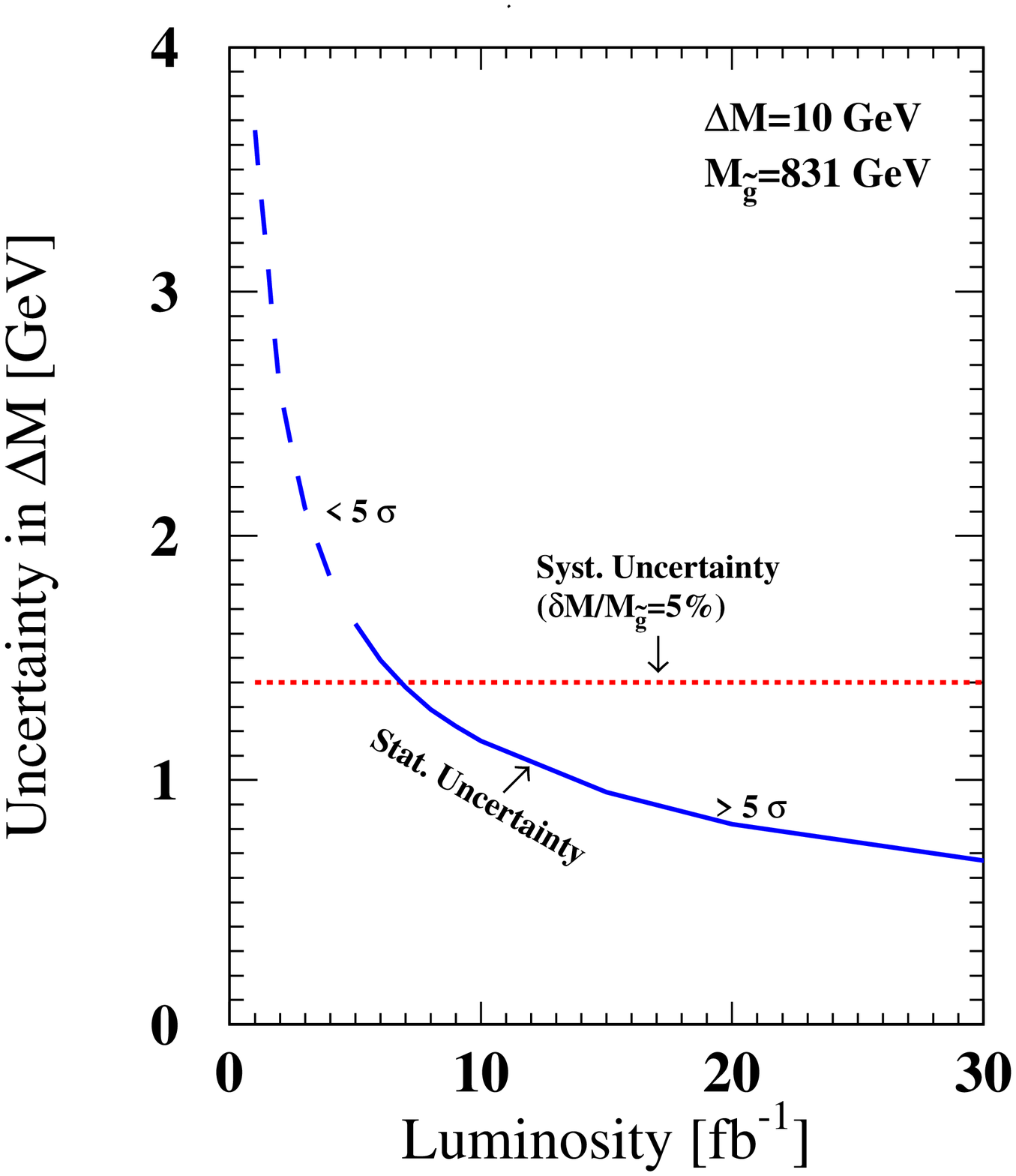} }
\caption{The uncertainty
in the $\dM$ measurement as a function of luminosity. The solid
(dashed) curve indicates where this analysis  has more than (less
than) 5$\sigma$ significance in OS$-$LS counts and shows the expected
statistical uncertainty. We also show the systematic uncertainty
(horizontal dotted line) in $\dM$ due to a gluino mass uncertainty  of
$\pm$5\%.
 }
\label{fig:uncertainty_on_dM}
\end{figure}

For the cases of $\dM$ = 6 and 10 GeV, with 10 \invfb\ of data, we
estimate that we can determine the $\dM$ values to be
\begin{equation}
\dM =
\begin{cases}
\begin{array}{rlll}
  6 & \pm 1.1 & ^{+1.0}_{-0.9} & \gev \\
 10 & \pm 1.2 & ^{+1.4}_{-1.2} & \gev
\end{array}
\end{cases}
\nonumber
\end{equation}
\noindent where the first uncertainty is due to the statistics and the
second due to a gluino mass variation of 5\% from the nominal value of
831 \gev. If one assumes a  gluino mass uncertainty of 10\%, the
second uncertainty is almost doubled to $^{+2.2}_{-1.7}$ and
$^{+3.2}_{-2.3}$ for the 6 and 10 GeV cases, respectively.
Figure~\ref{fig:uncertainty_on_dM} illustrates both uncertainties for
the $\dM$ = 10 GeV case as a function of luminosity.

\section{Conclusion}

 In mSUGRA models, the $\stauone$-$\schionezero$ coannihilation occurs
in a large region of parameter space allowed by the relic density
constraint. The characteristic feature of this coannihilation region
is a small mass difference ($\dM$) between the $\stauone$ and the
$\schionezero$. The small mass difference produces  final states
containing low energy $\tau$'s from $\schitwozero \rightarrow \tau
\stauone \rightarrow \tau\tau\schionezero$. We have demonstrated that
if LHC experiments reconstruct/identify $\tau$'s  with $\pt >$ 20 GeV
with an efficiency in the 50\% range, we could establish the signal
for this  coannihilation region using a sample of \met + 2j +
2$\tau$ events. Counting the number of OS$-$LS di-tau pairs in
conjunction with the observation of a mass peak below the expected end
points would establish the signal  for the case of $M_{\gluino} \simeq
830$~GeV with as few as 3-10~\invfb\ of data. We expect to require a
larger luminosity as $M_{\gluino}$ increases. We note that although we
have assumed that the jet $\rightarrow \tau$ fake probability was 1\%,
our results appear to be insensitive to the fake rate. Using the
di-tau mass measurement and assuming the gluino mass is already
measured, we have further shown that a measurement of $\dM$ is
possible even with 10 \invfb of data. For $\dM$ = 10 GeV, the
statistical uncertainty is 12\%, with a 14\% additional uncertainty
due to an assumed  5\% uncertainty in the gluino mass. Again, the
results are insensitive to the fake rate.

We note that the signal of \met + 1jet + 3$\tau$ also
 occurs
at a reduced rate from gaugino pairs, but with lower backgrounds.
However, it can provide a complementary signal that could help
solidify the discovery of the coannihilation signal at the LHC.

The coannihilation region is also present in most non-universal SUGRA
models, and an analysis similar to the one here for mSUGRA could be
performed for these. The main requirements for an observable signal
would be that the $\schitwozero$ decays predominantly to $\tau$'s, and
that the decay of the parent gluinos and squarks to $\schitwozero$ is
not suppressed.

\section*{Acknowledgments}
We thank Kamal Benslama for inspiring us
to use final states with two $\tau$'s
and for technical help throughout this analysis.
We also thank Y. Santoso for useful discussions.
This work is supported in part by a DOE Grant
DE-FG02-95ER40917, a NSF Grant PHY-0101015, and in part by NSERC of
Canada.


\begin{thebibliography}{99}

\bibitem{sugra01}
D.Z. Freedman, P. Van Niewenhuisen, and S. Ferrara,
\Journal{\PRD}{13}{3214}{1976}. 

\bibitem{sugra02}
S. Deser and B. Zumino,
\Journal{\PL}{65B}{369}{1976}. 

\bibitem{sugra1}
A.H. Chamseddine, R. Arnowitt, and P. Nath,
\Journal{\PRL}{49}{970}{1982}. 

\bibitem{sugra2}
R.~Barbieri, S.~Ferrara, and C.~A.~Savoy,
\Journal{\PLB}{119}{343}{1982}; L. Hall, J. Lykken, and S. Weinberg,
\Journal{\PRD}{27}{2359}{1983}; P. Nath, R. Arnowitt, and A.H.
Chamseddine,
\Journal{\NPB}{227}{121}{1983}.

\bibitem{nilles}
For a review, see P. Nilles,
\Journal{\PRep}{110}{1}{1984}. 


\bibitem{darkrv}
J. Ellis, K. Olive, Y. Santoso, and V. Spanos,
\Journal{\PLB}{565}{176}{2003};
R. Arnowitt, B. Dutta, and B. Hu,
hep-ph/0310103;
H. Baer, C. Balazs, A. Belyaev, T. Krupovnickas, and
X. Tata, \Journal{JHEP~}{0306}{054}{2003};
B. Lahanas and D.V. Nanopoulos, \Journal{\PLB}{568}{55}{2003};
U. Chattopadhyay, A. Corsetti, and P. Nath,
\Journal{\PRD}{68}{035005}{2003};
E. Baltz and  P. Gondolo, JHEP {\bf 0410} (2004) 052;
A. Djouadi, M. Drees, and J-L. Kneur, hep-ph/0602001.

\bibitem{sp}
WMAP Collaboration, D.N Spergel \etal,
Astrophys. J. Suppl. {\bf 148} (2003) 175.

\bibitem{Hinchliffe_Paige_2000}
I. Hinchlife and F.E. Paige,
\Journal{\PRD}{61}{095011}{2000}. 

\bibitem{Heinemann}
F.~Heinemann, hep-ex/0406056 (2004).

\bibitem{ilc_KADKstudy}
  V.~Khotilovich, R.~Arnowitt, B.~Dutta, and T.~Kamon,
\Journal{\PLB}{618}{182}{2005}.

\bibitem{higgs1}
ALEPH, DELPHI, L3, OPAL Collaborations,
G. Abbiendi, \etal\
(The LEP Working Group for Higgs Boson Searches),
 \Journal{\PLB}{565}{61}{2003}.

\bibitem{bsgamma}
M. Alam \etal, \Journal{\PRL}{74}{2885}{1995}. 

\bibitem{aleph}
Particle Data Group,
S. Eidelman \etal, \Journal{\PLB}{592}{1}{2004}.

\bibitem{BNL}
Muon $g-2$ Collaboration, G. Bennett \etal,
\Journal{\PRL}{92}{161802}{2004}. 

\bibitem{dav}
M. Davier, hep-ex/0312065.

\bibitem{hag}
K. Hagiwara, A. Martin, D. Nomura, and T. Teubner,
\Journal{\PRD}{69}{093003}{2004}. 

\bibitem{isajet}
F. Paige, S. Protopescu, H. Baer, and X. Tata, hep-ph/0312045.
We use~\isajet\ version 7.63.

\bibitem{atlfast}
{\tt ATLFAST} is the ATLAS fast simulation package, where
the simulation is performed by smearing the MC truth information directly with resolutions measured in full simulation studies.
We use version 7.0.2
(see http://www.hep.ucl.ac.uk/atlas/atlfast/).

\bibitem{3tau_paper}
R. Arnowitt, A. Aurisaro, B. Dutta, T. Kamon, N. Kolev,
P. Simeon, D. Toback, and P. Wagner,
in preparation.

\bibitem{cdfprl_himass_ditau}
CDF Collaboration, D. Acosta \etal, \Journal{\PRL}{95}{131801}{2005}.
The jet$\rightarrow\tau$ fake probability ($\faketau$) is measured to
be 1.1\% ($E_{\rm jet} \simeq$ 20 GeV) down to 0.2\% ($E_{\rm jet}\
\gtsim$ 100 GeV) as a function of the jet energy. We smply assume a
constant value (= 1\%) for our simulation study.

\bibitem{LHCstau_parameterization} In our simulation, we obtain
$N(x, y) = \left(-8.19 + 1.95\ x \right)
\cdot y^{-4.62}$,
$M_{\tau\tau}^{\rm peak}(x, y)  =
\left(30.3 + 1.96 x -0.0355 x^2  \right) \cdot
\left( 0.282 + 0.718\  y  \right)$, and
$\sigma_{M}(x) = 3.29 + 1.21\ x$.
Here $x$ = $\Delta M$ and $y$ = $M_{\gluino} / (831.2\ \gev)$.

\end{thebibliography}
\end{document}